\documentclass[12pt]{iopart}
\usepackage{graphicx}
\usepackage{iopams}

\newcommand{\bea}{\begin{eqnarray}}
\newcommand{\eea}{\end{eqnarray}}
\newcommand{\beq}{\begin{equation}}
\newcommand{\eeq}{\end{equation}}
\newcommand{\KMS}{\rm km\,s^{-1}}

\begin{document}

\def\fun#1#2{\lower3.6pt\vbox{\baselineskip0pt\lineskip.9pt
  \ialign{$\mathsurround=0pt#1\hfil##\hfil$\crcr#2\crcr\sim\crcr}}}
\def\lap{\mathrel{\mathpalette\fun <}}
\def\gap{\mathrel{\mathpalette\fun >}}
\def\kms{{\rm km\ s}^{-1}}
\def\vk{V_{\rm recoil}}

\title{Remnant masses, Spins and recoils from the Merger of Generic Black-Hole
Binaries}

\author{Carlos O. Lousto,
Manuela Campanelli,
Yosef Zlochower,
and Hiroyuki Nakano}

\address{Center for Computational Relativity and Gravitation,\\
and School of Mathematical Sciences, Rochester Institute of
Technology, 85 Lomb Memorial Drive, Rochester, New York 14623}

\begin{abstract}
We obtain empirical formulae for the final remnant black hole mass,
spin, and recoil velocity from merging black-hole binaries with
arbitrary mass ratios and spins. Our formulae are based on the mass
ratio and spin dependence of the post-Newtonian expressions for the
instantaneous radiated energy, linear momentum, and angular momentum,
as well as the ISCO binding energy and angular momentum. The
relative weight between the different terms is fixed by
amplitude parameters chosen through a least-squares fit of
recently available fully nonlinear numerical simulations.  These
formulae can be used for statistical studies of N-body simulations of
galaxy cores and clusters, and the cosmological growth of supermassive
black holes. As an example, we use these formulae to obtain a universal spin
magnitude distribution of merged black holes and recoil velocity
distributions for dry and hot/cold wet mergers. 
We also revisit the long term orbital precession and resonances and discuss
how they affect spin distributions before the merging regime.
\end{abstract}

\maketitle

\section{Introduction}\label{sec:Introduction}

Black holes at the centers of galaxies and globular clusters
significantly impact the dynamical evolution of these astronomical
structures. Of particular importance to the dynamics are the
black-hole (BH) mass, spin, and location (if off-center); properties that
can significantly change following a BH merger.  When two galaxies
merge, an event that is expected to occur several times during a
galaxy's evolution, the supermassive BH at their centers form a
black-hole binary (BHB) that eventually inspirals and merges.
Similarly, intermediate-mass BHs in globular clusters can form tight
BHBs that inspiral and merge.
Consequently, accurate models for the mass, spin, and gravitational
recoil of the  merger remnants of BHBs are of great
astrophysical interest. However, these models require accurate
simulations of merging BHs, a problem in the highly-nonlinear regime,
that only recently  became feasible due to breakthroughs in Numerical
Relativity~\cite{Pretorius:2005gq, Campanelli:2005dd, Baker:2005vv}.

The first attempts at modeling the remnant BH of BHB mergers using
fully nonlinear simulations utilized the `Lazarus method'
\cite{Baker:2003ds}, which combined short-term numerical simulations
of BHBs, just prior to merger,
with perturbative calculations.  With the advent of the `moving
punctures'~\cite{Campanelli:2005dd, Baker:2005vv} and
generalized harmonic~\cite{Pretorius:2005gq} approaches, it became
possible to accurately model merging BHBs from inspiral through merger
and ringdown using fully-nonlinear numerical simulations. 
As a result of these breakthroughs, NR groups from around the world
have been able to develop heuristic models for the properties of
remnant BHs as a function of the orbital and intrinsic BH parameters of the
binary (at-least for part of the parameter space).

The initial attempts at modeling the properties of remnant BH
 focused on the mass
and spin using {\em ad hoc} interpolation formulae.
 In \cite{Campanelli:2006uy, Campanelli:2006fg,
Campanelli:2006fy} we studied equal-mass, spinning BHBs, where the
individual BH spins were aligned and counter-aligned with the orbital
angular momentum, using fully nonlinear numerical calculations. We
found a simple quadratic polynomial relating the final mass and
spin of the remnant with the spins of the individual BHs. This
scenario was later revisited in \cite{Rezzolla:2007xa,
Rezzolla:2007rd}, and in \cite{Rezzolla:2007rz}  the authors
generalized the formula for the remnant spin (by assuming that the
angular momentum is only radiated along the orbital axis, and
neglecting the energy loss) in order to model arbitrary BH
configurations (these assumptions were relaxed in a follow-up
paper~\cite{Barausse:2009uz}).  A generic formulae for the final spin
was proposed in~\cite{Tichy:2008du} based on simulations with aligned
and non-aligned spins.  A more comprehensive approach, using a generic
Taylor expansion consistent with the the physical symmetries of the
problem, and with parameters chosen by a least-squares fit to many
simulations, was developed in \cite{Boyle:2007ru}.  All of these models
used low-order polynomial interpolation functions to predict the
remnant mass and spin as a function of the individual BH masses and
spins.  On the other hand, in \cite{Buonanno:2007sv}, a different
approach, based on approximate analytic models for the merger,  was used.
Here the authors extended the particle limit
approximation for the radiated mass and angular momentum to the
comparable-mass regime; ignoring effects of post-ISCO (Innermost
Stable Circular Orbit) gravitational radiation.  This approach was
further improved in~\cite{Kesden:2008ga} by taking binding energies
into account.  All of these approaches show a certain degree of
agreement with the remnant masses and spins obtained in the few dozen
fully nonlinear numerical simulations available, but significant
uncertainties concerning accuracy outside this range of the parameter
space remain.  Here we propose a set of formulae that incorporate the
benefits of both approaches and regimes in a unified way; using
analytic techniques to develop empirical models with free parameters
determined by numerical results.

Due to its significant impact on astrophysics, the modeling of the
remnant recoil followed an independent path,
particularly since the discovery \cite{Campanelli:2007ew,
Campanelli:2007cga}
that the spins of the black holes play a crucial role in producing
recoils of up to $4000\ \KMS$.
The realization that the merger of BHBs can produce recoil
velocities that allow the remnant to escape from
major galaxies led to numerous theoretical and observational efforts
to find traces of this phenomenon. Several studies made predictions of
specific observational features of recoiling supermassive black holes in the
cores of galaxies in the electromagnetic spectrum \cite{Haiman:2008zy,
Shields:2008va, Lippai:2008fx, Shields:2007ca, Komossa:2008ye,
Bonning:2007vt, Loeb:2007wz}. Notably, there began to appear
observations indicating the possibility of detection of such effects
\cite{Komossa:2008qd, Strateva:2008wt, Shields:2009jf}, and although alternative
explanations are possible \cite{Heckman:2008en, Shields:2008kn,
Bogdanovic:2008uz},  there is still the exciting possibility that these
observations can lead to the first confirmation of a prediction of
General Relativity in the highly-dynamical, strong-field regime.

This paper is organized as follows.
In Sec.~\ref{sec:recoil} we describe our empirical formula for the
remnant gravitational recoil and provide the leading coefficients for
this formula. In Sec.~\ref{sec:Mass} we describe our formula for the
final remnant mass, while in Sec.~\ref{sec:Spin} we describe the
formula for the final remnant spins. We provide fits to the constants
in the remnant mass and spin formula in Sec.~\ref{sec:fitting}.
In Sec.~\ref{sec:Inspiral} we revisit the gravitational
alignment and antialignment mechanisms for long term inspiral orbits, and
discuss the consequences and applications of our formulae in
Sec.~\ref{sec:Discussion}.

\section{Remnant Recoil Velocities}\label{sec:recoil}

In our approach to the recoil problem \cite{Campanelli:2007ew,
Campanelli:2007cga} we used post-Newtonian (PN) theory as a guide
to model the recoil dependence on the physical parameters of the
progenitor BHB (See Eqs.
(3.31) in \cite{Kidder:1995zr}), while arguing that only full
numerical simulations can produce the correct amplitude of the effect
(see Eq.~(\ref{eq:Pempirical}) below).
For example, in the instantaneous radiated linear momentum,
 there are terms of the form
\begin{equation}
\frac {d \vec P}{dt} = \cdots +   \frac{\eta^2}{1+q} \left[\vec F(\vec r, \vec v)\cdot
\left(\vec\alpha_2 - q \vec\alpha_1 \right)\right] \hat{\vec L},
\end{equation}
where $\hat {\vec L}$ is the unit vector pointing along the instantaneous
orbital angular momentum, $\vec F(\vec r, \vec v)$ is a vector in the
orbital plane that is only a function of the orbital position and its
time derivative, $q=m_1/m_2 \leq 1$ is the mass ratio, 
$\eta=q/(1+q)^2$ is the symmetric mass ratio, and 
$\vec \alpha_i = \vec{S_i}/m_i^2$ is the intrinsic spin on black hole $i$.
We incorporate this term by adding a term 
\begin{equation}
\vec{V}_{\rm recoil} = \cdots + \left(K\frac{\eta^2}{(1+q)}\left[
\left|\alpha_2^\perp-q\alpha_1^\perp\right|
\cos(\Theta_\Delta-\Theta_0)\right]\right)\hat{\vec L}
\end{equation}
 to our fitting
formula for the recoil velocity (see Eq.~(\ref{eq:Pempirical}) below),
 where the fitting constants $K$ and $\Theta_0$ approximate the
 net effect
of the dynamics of this term during the last few $M$ of the rapid plunge
 (where most of
the recoil is generated) and $\Theta_\Delta$ is the angle that
$\vec \Delta = M^2(\vec \alpha_2 - q \vec \alpha_1)/(1+q)$ 
where $M=m_1+m_2$, makes with the infall direction at merger.
Our heuristic formula describing the recoil
velocity of BHB remnants 
 was theoretically verified in
several ways. In~\cite{Campanelli:2007cga} we confirmed
the sinusoidal dependence [$\cos \Theta_\Delta$ in Eq.~(\ref{eq:Pempirical})]
of the recoil on the direction  of the
in-plane spin for the so-called `superkick' configurations, 
a result that was  reproduced in~\cite{Brugmann:2007zj}
for binaries with different initial separations. While
in \cite{Herrmann:2007ex} the authors verified the decomposition of the
spin-dependence of the recoil into spin components
perpendicular and parallel to the orbital plane. Similarly, in
\cite{Pollney:2007ss} the authors determined that the quadratic-in-spin
corrections to the in-plane recoil velocity are less than 
$5\%$ of the total recoil.
Recently in \cite{Lousto:2008dn} we confirmed the leading $\eta^2$
(where $\eta$ is symmetric mass ratio) dependence of the large recoils
 out of the orbital plane (see also \cite{Baker:2008md}).

Here we augment our original empirical formula with
subleading terms, higher order in the mass ratio,
and include a new term linear in the total spin, motivated by higher
order post-Newtonian computations~\cite{Racine:2008kj},
\begin{eqnarray}\label{eq:Pempirical}
\fl
\vec{V}_{\rm recoil}(q,\vec\alpha)=v_m\,\hat{e}_1+
v_\perp(\cos\xi\,\hat{e}_1+\sin\xi\,\hat{e}_2)+v_\|\,\hat{n}_\|,\nonumber\\
v_m=A\frac{\eta^2(1-q)}{(1+q)}\left[1+B\,\eta\right],\nonumber\\
v_\perp=H\frac{\eta^2}{(1+q)}\left[
(1+B_H\,\eta)\,(\alpha_2^\|-q\alpha_1^\|)
+\,H_S\,\frac{(1-q)}{(1+q)^2}\,(\alpha_2^\|+q^2\alpha_1^\|)\right],\nonumber\\
v_\|=K\frac{\eta^2}{(1+q)}\Bigg[
(1+B_K\,\eta)
\left|\alpha_2^\perp-q\alpha_1^\perp\right|
\cos(\Theta_\Delta-\Theta_0)\nonumber\\
\quad+\,K_S\,\frac{(1-q)}{(1+q)^2}\,\left|\alpha_2^\perp+q^2\alpha_1^\perp\right|
\cos(\Theta_S-\Theta_1)\Bigg],
\end{eqnarray}
where 
 the index $\perp$ and $\|$ refer to
perpendicular and parallel to the orbital angular momentum respectively 
and $\hat{n}_\|=\hat{\vec L}$. 
$\hat{e}_1,\hat{e}_2$ are
orthogonal unit vectors in the orbital plane, and $\xi$ measures the
angle between the unequal mass and spin contribution to the recoil
velocity in the orbital plane. The new constants $H_S$ and $K_S$ can be 
determined from new generic BHB simulations as the data become
 available. The angles, $\Theta_\Delta$ and $\Theta_S$, are
the angles
between the in-plane component of $\vec \Delta = M (\vec S_2/m_2 - \vec
S_1/m_1)$ or $\vec S=\vec S_1+\vec S_2$
and the infall direction at merger. Phases $\Theta_0$ and $\Theta_1$ depend
on the initial separation of the holes for quasicircular orbits.  
A crucial observation is that the dominant contribution to the recoil
is generated near the time of formation of the common horizon of the
merging black holes (See, for instance Fig. 6 in
~\cite{Lousto:2007db}).  The formula~(\ref{eq:Pempirical}) above
describing the recoil applies at this moment (or averaged coefficients
around this maximum generation of recoil), and has proven to represent
 the distribution of velocities with sufficient accuracy for astrophysical
applications.  

The most recent estimates for the above parameters can be found in 
\cite{Lousto:2008dn} and references therein. The current best
estimates are: $A = 1.2\times 10^{4}\
\kms$, $B = -0.93$, $H = (6.9\pm0.5)\times 10^{3}\ \kms$,
$K=(6.0\pm0.1)\times 10^4\ \kms$, and $\xi \sim 145^\circ$.  Note that
we can use the data from~\cite{Lousto:2008dn} to obtain
$K=(6.072\pm0.065)\times 10^4\ \KMS$, if we assume that $B_{K}$ and
$K_{S}$ are negligible. 
Finally, if we fit the data to find $K$ and $K_{S}$
simultaneously we obtain $K=(6.20\pm0.12)\times 10^{4}\ \KMS$ and
$K_{S} = -0.056\pm0.041$, where we made the additional assumption that
$\Theta_0=\Theta_1$ (since $\vec S=\vec \Delta$ for these runs).
An attempt to fit $K$, $K_S$, $B_K$ simultaneously 
does not produce robust results with currently available data 
(one of the reasons for this is that different values of $K$ and $B_K$
produce very similar predicted recoil velocities over the range $0.1 \leq q
\leq 1$).
Note that the values for the
dominant $K$ term are reasonably insensitive to the different choices
for the fits, while finding the subleading terms require additional
 runs and higher accuracy.

The above equation (\ref{eq:Pempirical}) contains all the expected 
linear terms in the spin, and includes ten fitting parameters.
Based on the works \cite{Racine:2008kj} one could add quadratic terms,
and this will be published elsewhere by the authors.

From a practical point of view, for statistical simulations of BHB
mergers, where the directions of the spins and infall direction is not known,
 one should take a uniform distribution  for the in plane-components
of $\hat \alpha_1$ and
$\hat \alpha_2$ over all possible angles, define $\Theta_S$ and 
$\Theta_\Delta$ with respect to a fixed arbitrary in-plane 
vector (say $\hat x$), and take $\Theta_0=0$. The
resulting distribution of recoil velocities will be independent of the
choice of the arbitrary in-plane vector (but will depend
weakly on $\Theta_1$). If we ignore the subleading $K_S$ correction,
then $\Theta_1$ will not enter the recoil calculation. It's effects
can be incorporated by including the $K_S$ term and 
averaging over all possible values of
$\Theta_1$.

\section{Remnant Mass}\label{sec:Mass}

Motivated by the success of the empirical formula for the recoil, we
propose a new empirical formula for the total radiated energy based on
the post-Newtonian equations for the instantaneous radiated
energy (see Eqs.~(3.25) in \cite{Kidder:1995zr}, and for the
quadratic terms in the spin see Eq.~(5.4) in \cite{Racine:2008kj}):
\begin{eqnarray}\label{Eempirical}
\fl
\delta M/M = \eta\,\tilde{E}_{ISCO}
+E_2\eta^2+E_3\eta^3
\nonumber\\
+\frac{\eta^2}{(1+q)^2}\Bigg\{E_S\,(\alpha_2^\|+q^2\,\alpha_1^\|)
+E_\Delta\,(1-q)\,(\alpha_2^\|-q\,\alpha_1^\|)+ E_A\,|\vec\alpha_2+q\vec\alpha_1|^2\nonumber\\
+E_B\,|\alpha_2^\perp+q\alpha_1^\perp|^2
\left(\cos^2(\Theta_{+}-\Theta_2)+E_C\right)+ E_D\,|\vec\alpha_2-q\vec\alpha_1|^2\nonumber\\
+E_E\,|\alpha_2^\perp-q\alpha_1^\perp|^2
\left(\cos^2(\Theta_{-}-\Theta_3)+E_F\right)\Bigg\},
\end{eqnarray}
where 
$\Theta_{\pm}$
are the angles that  $\vec\Delta_\pm=M(\vec S_2/m_2\pm\vec S_1/m_1)$ make 
with the radial direction
during the final plunge and merger (for comparable-mass BHs, a 
sizable fraction of the 
radiation is emitted during this final plunge, see for instance Fig.~6 in
\cite{Lousto:2007db}). Phases $\Theta_{2,3}$ are parameters that give the
angle of maximum radiation for these terms, and depend on the initial 
separation and parameters of the binary at the beginning of the numerical
simulation. 

In addition to the terms arising from the
instantaneous radiated energy, which gives 12 fitting parameters,
we also included terms associated with the secular loss of energy in the
inspiral period from essentially infinite separation down to the plunge.
In order to model this contribution we adopted the effective one body
form \cite{Damour:2008qf} supplemented by the
 $\eta^2$ effects from self force
calculations \cite{Barack:2009ey} and 2PN effects of the 
spins (see Eq.~(4.6) in \cite{Kidder:1995zr}), to obtain
\begin{eqnarray}\label{EISCO}
\fl
\tilde{E}_{ISCO}\approx
(1-\sqrt{8}/3)+0.103803\eta \nonumber \\
+
\frac{1}{36\sqrt{3}(1+q)^2}
\left[q(1+2q)\alpha^\|_1+(2+q)\alpha^\|_2\right]\nonumber\\
-\frac{5}{324\sqrt{2}(1+q)^2}\left[
\vec\alpha^2_2-3(\alpha_2^\|)^2
-2q(\vec\alpha_1\cdot\vec\alpha_2-3\alpha_1^\|\alpha_2^\|)
+q^2(\vec\alpha^2_1-3(\alpha_1^\|)^2) \right] \nonumber \\
+ {\cal O}(\alpha^3).
\end{eqnarray}
The above expression only includes quadratic-in-spin terms for
compactness,  hence it is expected to produce reliable results for 
intrinsic spin magnitudes
$\alpha_i<0.8$ (because the binding energy is a very steep function of
$\alpha$ for $\alpha > 0.8$ and the quadratic expressions above are no
longer appropriate).
 Note that we used the full expressions from
\cite{Damour:2008qf} to obtain our fitting parameters.
Here we fit the leading-order parameters using available
data, and as new data become available, we expect to be able fit the
remaining parameters. 

\section{Remnant Spin}\label{sec:Spin}

In an analogous way, we propose an empirical formula for the
final remnant spin 
based on the post-Newtonian equations for the radiated
angular momentum and the angular
momentum of a circular binary at close separations
(see Eqs.~(3.28) and (4.7) in \cite{Kidder:1995zr}), 
\begin{eqnarray}\label{Jempirical}
\fl
\vec{\alpha}_{\rm final} = \left(1-\delta M/M\right)^{-2}\Big\{\eta\tilde{\vec{J}}_{ISCO}+
\left(J_2\eta^2+J_3\eta^3\right)\hat{n}_\|\nonumber\\
+\frac{\eta^2}{(1+q)^2}\left(\left[J_A\,(\alpha_2^\|+q^2\,\alpha_1^\|)
+J_B\,(1-q)\,(\alpha_2^\|-q\,\alpha_1^\|)\right]\hat{n}_\| \right.\nonumber\\
\left.+(1-q)\,|\vec\alpha_2^\perp-q\,\vec\alpha_1^\perp|
\sqrt{J_\Delta\cos[2(\Theta_\Delta-\Theta_4)]+J_{M\Delta}}\,\hat{n}_\perp \right.\nonumber\\
\left.+|\vec\alpha_2^\perp+q^2\,\vec\alpha_1^\perp|
\,\sqrt{J_S\cos[2(\Theta_S-\Theta_5)]+J_{MS}}\,\hat{n}_\perp
\right)\Big\}.
\end{eqnarray}
Note that, even at linear order, there are important contributions 
of generic spinning black holes producing
radiation in directions off the orbital axis that do not vanish in the
equal-mass or zero-total-spin cases.
The above formula can be augmented by quadratic-in-the-spins
terms~\cite{Racine:2008kj,LNZ}
of a form similar to the terms added to the radiated
energy formula (\ref{Eempirical}). However, these terms are less
significant for modeling the final spin 
(see, for instance Fig.~21 of \cite{Campanelli:2006fy}).

Again, we use the effective one body resummation
form \cite{Damour:2008qf}, supplemented with the
$\eta^2$ effects from self force
calculations \cite{Barack:2009ey} and the 2PN effects of the 
spins (see Eq.~(4.7) in \cite{Kidder:1995zr}), to obtain
\begin{eqnarray}\label{JISCO}
\fl
\tilde{\vec{J}}_{ISCO}\approx\Bigg\{2\sqrt{3}
-1.5255862\eta 
-\frac{1}{9\sqrt{2}(1+q)^2}
\left[q(7+8q)\alpha^\|_1+(8+7q)\alpha^\|_2\right]\nonumber\\
+\frac{2}{9\sqrt{3}(1+q)^2}\left[
\vec\alpha^2_2-3(\alpha_2^\|)^2 
-2q(\vec\alpha_1\cdot\vec\alpha_2-3\alpha_1^\|\alpha_2^\|)
+q^2(\vec\alpha^2_1-3(\alpha_1^\|)^2)
\right]\bigg\}\hat{n}^\|\nonumber\\
-\frac{1}{9\sqrt{2}(1+q)^2}
\left[q(1+4q)\vec\alpha_1+(4+q)\vec\alpha_2\right]
+\frac{1}{\eta}\frac{(\vec\alpha_2+q^2\vec\alpha_1)}{(1+q)^2} 
+{\cal O}(\alpha^3).
\end{eqnarray}
This expression represents a quadratic expansion in the spin-dependence,
hence we expect to produce reliable results for intrinsic spin magnitudes
$\alpha_i<0.8$ (hence $\alpha_{\rm final} < 0.9$).

\section{Determination of fitting parameters}\label{sec:fitting}

Here we show how results from current full numerical simulations can
be used to determine the fitting constants in the equations for
the final remnant mass and spins of a BHB merger.
This procedure can be repeated and extended as we have access
to new runs and can also help in designing new 
simulations to optimally determine all fitting constants and
better cover the 7-dimensional physical parameter space of
BHBs. We used Mathematica's LinearRegression and
 NonLinearRegress functions to find the fitting parameters and
estimate the errors in the parameters. Our method for finding the
fitting parameters was to first fit to simulations with symmetries
that caused most terms to vanish in order to fit to as few parameters
at a time as possible. Then, after fixing the parameters we found in earlier
fits, we fit to simulations with less symmetry to obtain other
parameters. For example, we first find $E_2$ and $E_3$, and then
using these values, fit additional data to obtain $E_S$, etc.

{\it Energy radiated: }
For the non-spinning case, we fit the data from 8 simulations 
found in
\cite{Berti:2007fi, Gonzalez:2008bi} (see also
\cite{Buonanno:2007pf}). Here we fit $E_{\rm Rad}$
versus $\eta$, where $E_{\rm Rad}$ is the total radiated energy for a
given configuration minus the binding energy of the initial
configuration (where the binding energy is negative). We calculate the
binding energy using the 3PN accurate expressions given in
\cite{Blanchet:2000nv}.
A fit of the resulting data gives  $E_2 = 0.341\pm0.014$ and $E_3 =
0.522\pm0.062$.
In order to estimate $E_{S}$, $E_{\Delta}$, $E_A$, and $E_D$, We use
the remnant masses from 13 simulations
for spinning BHBs with spins aligned
with the orbital angular momentum given in \cite{Berti:2007nw,
Hannam:2007wf} (see also
\cite{Campanelli:2006uy}), and find
$E_{S} = 0.673\pm0.035$, $E_{\Delta} = -0.36\pm0.37$,  $E_{A} =
-0.014\pm0.021$, and $E_{D} = 0.26\pm0.44$. 
These large uncertainties in the fitting parameters are due to the
effect of correcting for the binding energy in these simulations.
Finally, fits from the final remnant masses from 5 simulations
\cite{Campanelli:2007cga} yields $E_E=0.09594\pm0.00045$ and
fits from 5 equal-mass configurations in \cite{Lousto:2008dn}
yield $E_B = 0.045\pm0.010$. An
accurate fit to $E_D$ is not possible with the configurations available
in~\cite{Lousto:2008dn}. Note that our fits for $E_\Delta$, $E_A$, 
and $E_D$ are consistent with the parameters set to zero. This is due
to the fact that the errors introduced in renormalizing the data are
of the same order as the effects of these subleading terms.

{\it Angular momentum radiated: }
For the non-spinning case, we fit the data from 8 simulations in
\cite{Berti:2007fi, Gonzalez:2008bi}, and find
$J_2 = -2.81\pm0.11$ and $J_3 = 1.69\pm0.51$.
A fit to  $J_{A}$ and $J_{B}$ from 13 simulations 
in~\cite{Berti:2007nw, Hannam:2007wf} yields
$J_{A} = -2.97\pm0.26$ and $J_{B} = -1.73\pm0.80$. However,
we determined that the uncertainty in $J_{A}$ and $J_{B}$ is actually 
closer to $1.0$ by considering fits to the independent datasets
 in~\cite{Berti:2007nw}.

From the combined fit, we find that $2.42\% < \delta M/M <
9.45\%$ and $0.34 < \alpha_{\rm final} < 0.92$ for the equal-mass, aligned spin
scenario, in the region where the fit is valid ($|\alpha_{\rm final}| < 0.9$).

Finally, we note that much of the errors in the fitting parameters are
due to differences in the normalizations between the various runs.
Some authors choose normalize their simulations such that
$m_1 + m_2 = 1$, which approximates a binary that inspiraled from
infinity with an initial mass of 1, while others choose to normalize
their simulations such that the initial ADM mass is 1. In this latter
case we attempted to renormalize the results using the 3PN expression
for the binding energy. However, the errors introduced by
renormalizing data, or assuming that the ADM mass at infinite
separation is 1, introduces uncertainties in our fitting
parameters. This affects both $\delta M/M$ directly and
$\vec{\alpha}_{\rm final}$ indirectly through $\delta M/M$. Ideally we
would use a set of simulations with the same normalization and all
starting from the same initial orbital frequency.

From a practical point of view, for statistical simulations of BHB
mergers, where the infall direction and the directions of the spins at
merger are not known, one should take a uniform distribution  for the
in plane-components of $\hat \alpha_1$ and $\hat \alpha_2$ over all
possible angles, define the angles $\Theta_S$, $\Theta_\Delta$, and
$\Theta_+$ (note $\Theta_- = \Theta_\Delta$) with respect to a fixed
arbitrary in-plane vector (say $\hat x$), and take a uniform
distribution for the unknown angles $\Theta_{1,3,5}$. The angles
$\Theta_{0,2,4}$ can be set to zero, since the final distributions
will be independent of this choice (the distribution will only be a
weak function of the relative angles $\Theta_0-\Theta_1$, etc.).  The
resulting distributions will be independent of the choice of the
arbitrary in-plane vector (but will depend weakly on
$\Theta_{1,3,5}$). However, the angles $\Theta_{1,3,5}$ only appear in
subleading expressions and the uncertainties in the final
distributions of the spins, masses, and recoils should not be
significant for astrophysical applications.

\section{Inspiral phase}\label{sec:Inspiral}

One of the important application of our formulae is to study statistical
distributions of the final mass, spin and recoil of the remnant merged black
hole given an initial distribution of individual spins and mass ratios.
This kind of studies have been performed lately, see for instance \cite{Lousto:2009ka},
assuming initial random distribution of individual spin directions and magnitudes
as well as mass ratios. This choice was supported by the 
post-Newtonian studies \cite{Lousto:2009ka} that
in the (dry) inspiral phase, preliminary to the final merger we have
modeled in this paper, there is not an strong alignment of the spins with the orbital
angular momentum, as there would be if, for instance, we would have large
accretion of gas in the system (wet mergers).

The simulations in \cite{Lousto:2009ka} actually found that, gravitational
radiation induced precession of the orbital plane during the inspiral phase leads
to an small bias of the spins towards counter-alignment. These results were
the product of integrating 3.5PN equations of motion form separations $r=50M$
down to the merger regime around $r=5M$. It was point out by studying averaged
PN equations in the quasicircular orbits regime 
\cite{Schnittman:2004vq,Herrmann:2009mr,Kesden:2010yp}, 
that on longer time scales there are resonances that might affect the
distributions of spin directions by the time of merger.
Since these studies are complementary to those presented in 
\cite{Lousto:2009ka}, we will investigate this issue analytically at a lower PN order than
in the numerical studies of Ref.~\cite{Lousto:2009ka}, but retaining the radiation reaction effects 
on the orbital 
plane for consistency with the integration of the PN equations of motion
in the Hamiltonian formalism.

In terms of the notation and approach of Ref.~\cite{Lousto:2009ka}
we consider
\begin{eqnarray}
{(\hat {\vec{L}} \cdot \hat {\vec{S}}_i)}^\cdot 
&=& \frac{\dot {\vec{L}} \cdot {\vec{S}}_i}{|\vec{L}||{\vec{S}}_i|} 
+ \frac{\vec{L} \cdot \dot {\vec{S}}_i}{|\vec{L}||{\vec{S}}_i|}
- \frac{\vec{L} \cdot {\vec{S}}_i\,|\vec{L}|^\cdot}{|\vec{L}|^2|{\vec{S}}_i|} 
- \frac{\vec{L} \cdot {\vec{S}}_i\,|{\vec{S}}_i|^\cdot}{|\vec{L}||{\vec{S}}_i|^2} 
\,,
\end{eqnarray}
where we can set $|{\vec{S}}_i|^\cdot=0$ at this order of approximation. 
In \cite{Lousto:2009ka}  $\dot {\vec{S}}_i$ and the conservative part of $\dot {\vec{L}}$ terms did not contribute
due to the nature of the statistical studies performed in that paper. 
We hence focus on the dissipative part. 
\begin{eqnarray}
{(\hat {\vec{L}} \cdot \hat {\vec{S}}_i)}^\cdot_{\rm dis} 
&=& \frac{\dot {\vec{L}}_{\rm dis} \cdot {\vec{S}}_i}{|\vec{L}||{\vec{S}}_i|} 
- \frac{\vec{L} \cdot {\vec{S}}_i\,|\vec{L}|^\cdot_{\rm dis}}{|\vec{L}|^2|{\vec{S}}_i|} 
\,,
\end{eqnarray}
With the PN techniques described \cite{Lousto:2009ka} we find
\begin{eqnarray}
&&(\hat {\vec{L}} \cdot \hat {\vec{S}}_1)^{\cdot}_{\rm dis}  
= 
- \frac{8}{15} \, \frac{v_{\omega}^{11}}{M} 
\frac{q}{(1+q)^4} \, \frac{1}{|\vec{\alpha}_1|} 
\nonumber \\ && \qquad \times 
\biggl\{
q\,\left(61\,q + 48 \right) (\hat {\vec{P}} \cdot \vec{\alpha}_1)^2 
+ \left( 61 +48 \, q \right) 
(\hat {\vec{P}} \cdot \vec{\alpha}_1)(\hat {\vec{P}} \cdot \vec{\alpha}_2) 
\biggr\} 
\,, 
\label{eq:PNpredS1}
\\
&&(\hat {\vec{L}} \cdot \hat {\vec{S}}_2)^{\cdot}_{\rm dis}  
= 
- \frac{8}{15} \, \frac{v_{\omega}^{11}}{M} 
\frac{q}{(1+q)^4} \, \frac{1}{|\vec{\alpha}_2|} 
\nonumber \\ && \qquad \times 
\biggl\{
q\,\left(61\,q + 48 \right) (\hat {\vec{P}} \cdot \vec{\alpha}_1)
(\hat {\vec{P}} \cdot \vec{\alpha}_2) 
+ \left( 61 +48 \, q \right) (\hat {\vec{P}} \cdot \vec{\alpha}_2)^2 
\biggr\} 
\,. 
\label{eq:PNpredS2}
\end{eqnarray}
Note that this expressions are defined negative when averaged over
spin directions with only the squared terms 
$(\hat {\vec{P}} \cdot \vec{\alpha}_i)^2$ contributing.
By integrating them over time, we obtain similar results
to the expression for $(\hat {\vec{L}} \cdot \hat {\vec{S}})^{\cdot}_{\rm dis}$ 
in Eq (18) of \cite{Lousto:2009ka}, that lead us to the conclusion 
that distributions of spins show some bias towards counter-alignment with
respect to the orbital angular momentum. Note that the instantaneous
counteralignment mechanism acts at every radius, with increasing strength
for small separations, where the orbital velocity $v_\omega$ is large.

To investigate the small mass ratio limit, i.e. $q\to0$,
we compute the time integral (roughly speaking, multiply by 1/q) 
of Eq.~(\ref{eq:PNpredS2}), for instance.
We can then see that if the larger black hole's spin, $\vec S_2$, is 
initially randomly distributed in the limit $q \to 0$ it ends up with
some counteralignment. 
On the other hand, the smaller black hole's spin, $\vec S_1$, would 
remain to be 
random oriented as seen from the vanishing of the right hand side of
Eq.~(\ref{eq:PNpredS1}).

Note that the above equations do not use orbital averages since
the effect is particularly strong in the latest part of the inspiral,
when averages are not a good approximation. In the alternative regime,
when the inspiral motion is very slow, resonance orbits have been found
using orbit averaged descriptions 
\cite{Schnittman:2004vq,Herrmann:2009mr,Kesden:2010yp,Kesden:2010ji}.
These resonance orbits lead to alignment or antialignment of spins
if one starts from an initial aligned or antialigned large hole and
allow random orientations for the less massive one.
Note that if both spins are allowed to be chosen at random initially,
as we assumed in our computations,
then the resulting evolution leads to still random distributed spins.

The resonance mechanism is complementary to the mechanism we studied
in \cite{Lousto:2009ka}. The former takes place on very long time scales
compared to precession, while the later mechanism is quadratic in the
spins (as seen in Eqs. (\ref{eq:PNpredS1}) and (\ref{eq:PNpredS2}),
hence higher order.)
In order to quantify which of them is the predominant mechanism long term 
numerical integration
of the (non averaged) equations of motion is required.

\section{Discussion}\label{sec:Discussion}

In this paper, we provided a framework to describe the bulk properties
of the remnant of a BHB merger. Our framework is based on PN scaling
and fitting the results of full numerical simulations.  The new
formulae are physically motivated, incorporate the correct mass ratio
dependence, and account for the radiation of angular momentum both
parallel and perpendicular to the orbital angular momentum.  These
formulae have a symmetric dependence on the mass ratio and spins,
while still including the correct particle limit.  We also extended
the  successful recoil formula (\ref{eq:Pempirical}) by adding
nonleading terms that include all the linear dependence in the spins,
as well as  higher mass ratio powers.

Unlike in the formula for the remnant recoil case, the energy lost by
the binary during the inspiral phase is a non-trivial fraction of the
total radiated energy (and is, in fact, the dominant contribution in
the small mass ratio limit). We thus included both the instantaneous
radiative terms in  (\ref{Eempirical}) and the binding energy at the
ISCO Eq.~(\ref{EISCO}) (to take into account the secular loss of energy
from very large distances down into the merger and plunge regime).
Similarly, in order to model the final remnant spin, we need to take
into account  both the angular momentum of the system near the ISCO,
see Eq.~(\ref{JISCO}), and the subsequent loss of angular momentum in the
final plunge (which is particularly important in comparable-mass
mergers).

Using the fitted coefficients in the above formulae, we find that
for equal-mass, non-spinning binaries, the net energy radiated is
$5\%$ of the total mass and the final spin is $\alpha\approx0.69$,
both in good agreement with the most accurate full
numerical runs \cite{Scheel:2008rj}. For maximally spinning BHBs
with spin aligned and counteraligned  we estimate that quadratic corrections
lead to radiated energies between $10\%$ and $3\%$ respectively. As for
the magnitude of the remnant spin, the linear estimates are between
$0.97$ and $0.41$ respectively, with quadratic corrections slightly
reducing
those values. These results show that the cosmic censorship hypothesis
is obeyed (i.e.\ no naked singularities are formed) and are in good
agreement with earlier estimates \cite{Campanelli:2006fy}.


\begin{table}
\caption{The following parameters give the current best estimates for
the constants in Eqs.~(\ref{Eempirical})~and~(\ref{Jempirical}). These
parameters were used to generate the spin-magnitude distribution in
Fig.~\ref{fig:alphadist}.}
\label{tab:params}
\begin{center}
\begin{tabular}{lr|lr|lr|lr}
\hline
$E_2$ & $0.341$ & $E_3$ & $0.522$& $E_\Delta $ & $-0.3689$ & $E_A$ & $-0.0136$\\
$E_B$ & $ 0.045$ & $E_C$ & $0$ & $E_D$ & $0.2611$ & $E_E$ & $0.0959$ \\
$E_F$ & $ 0$ & & & &\\
$J_2$ & $-2.81$ & $J_3$ & $1.69$ & $J_A $ & $-2.9667$ & $J_B$ & $-1.7296$\\
$J_\Delta$ & $0$ & $J_M$ & $0$ & $J_{S}$ & $0$ & $J_{MS}$ & $0$\\
\hline
\end{tabular}
\end{center}
\end{table}

\begin{figure}
\begin{center}
\includegraphics[width=3.5in]{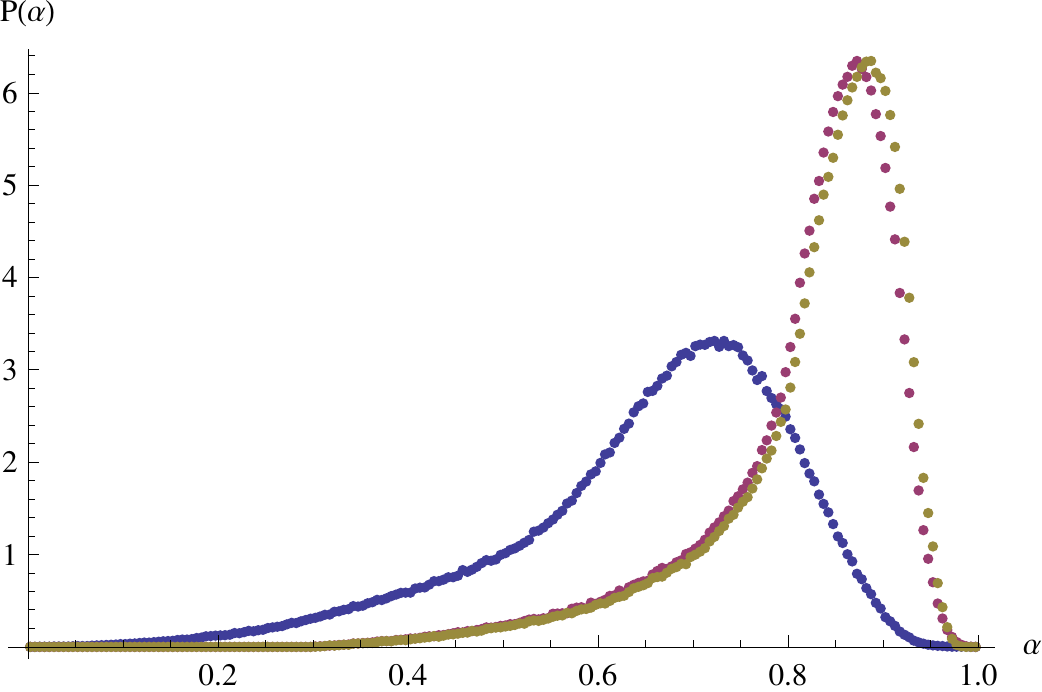}
\caption{The spin magnitude distribution for dry mergers. 
We plot the distributions of
spins $\alpha = S/m^2$ of the final remnant after many mergers. This
distribution does not change significantly following additional
mergers and peaks at $\alpha\approx0.73$. 
We also display the distributions representing wet mergers for
hot and cold accretion disks. They are highly peaked distributions
at around $\alpha\approx0.88$ and $\alpha\approx0.9$ respectively.  
}
\label{fig:alphadist}
\end{center}
\end{figure}

The set of formulae  (\ref{Eempirical}) and  (\ref{Jempirical}) with
the fitting constants determined as in the Sec.~\ref{sec:fitting} 
can be used to describe the final stage of BHB
mergers in theoretical, N-body, statistical studies in
astrophysics and cosmology \cite{Lousto:2009ka, O'Leary:2008mq, Blecha:2008mg,
Miller:2008yw, Kornreich:2008ca, Volonteri:2007dx, Gualandris:2007nm,
HolleyBockelmann:2007eh, Guedes:2008he, Volonteri:2007ax,
Schnittman:2007nb, Bogdanovic:2007hp, Schnittman:2007sn, Berti:2008af}
by choosing a distribution of the initial intrinsic physical
parameters of the binaries $(q, \vec{S}_1, \vec{S}_2)$ and mapping them to the
final distribution of recoil velocities, spins and masses
after the mergers.
As an example of such an application of the above formulae, we calculate the
expected distribution of spin magnitudes of astrophysical supermassive
and intermediate-mass BHs (which
are expected to have undergone several mergers).
To do this, we first
consider a set of $10^6$ binaries with uniform distributions of
mass ratio (from 0 to 1), uniform orientations of the spin
directions, and uniform distributions of spin magnitudes.
We then use our formulae (see Table~\ref{tab:params} for the values of
the constants that we used) to predict the spin-magnitude distribution 
of the merger
remnants and repeat the calculation, again with uniform distributions
in mass ratio and spin directions, but with this new spin-magnitude
distribution (see also \cite{Berti:2008af,
Tichy:2008du}). The resulting spin-distribution, after each subsequent
set of mergers, approaches a fixed distribution.
The spin distribution that results after ten generations of mergers 
is shown in Fig.~\ref{fig:alphadist}. The final results are insensitive to the initial
distribution and quickly converge, in a few generations of mergers, to the
displayed curve, which represents a universal distribution of the
intrinsic spin magnitudes (with a maximum near 0.7 and
mean in the range $(0.5, 0.8)$) 
of the remnant BHs of dry BHB mergers (when neglecting accretion).
In order to provide a simple analytical model for this distribution,
we fit it to the Kumaraswamy functional form
\cite{Jones:2009}
$f(x; a,b) = a b x^{a-1}{ (1-x^a)}^{b-1}$, and find
$a=5.91\pm0.04$, $b=5.33\pm0.07$. We choose this functional form
because it allows for a skewed distribution (and fits the results
better than a beta distribution), however,
the fit underestimates the probability for
producing small spins.

We have also considered the effect of wet mergers on the final spin
and recoil velocities. A first account of accretion effects during the
long inspiral phase of binary black holes have been given in \cite{Dotti:2009vz}
using the smoothed particle hydrodynamics approximation (SPH).
To evaluate the accretion effects on the statistical distributions,
according to \cite{Dotti:2009vz}, we have considered distributions that at merger
have $0.3\leq\alpha_i\leq0.9$ and orientations within 10 deg and 30 deg for
cold and hot accretion disks respectively. We have also assumed a
flat distribution in mass ratios in the region $0\leq q\leq1$.
The results for the final spin distributions are displayed in
 Fig.~\ref{fig:alphadist} and show the dramatic change in the spin
distributions due to accretion. Note that this accretion effects on
spin will be less important on black holes with masses larger than
$10^7M\odot$ \cite{Volonteri:2010hk}.

\begin{figure}
\begin{center}
\includegraphics[width=3.5in]{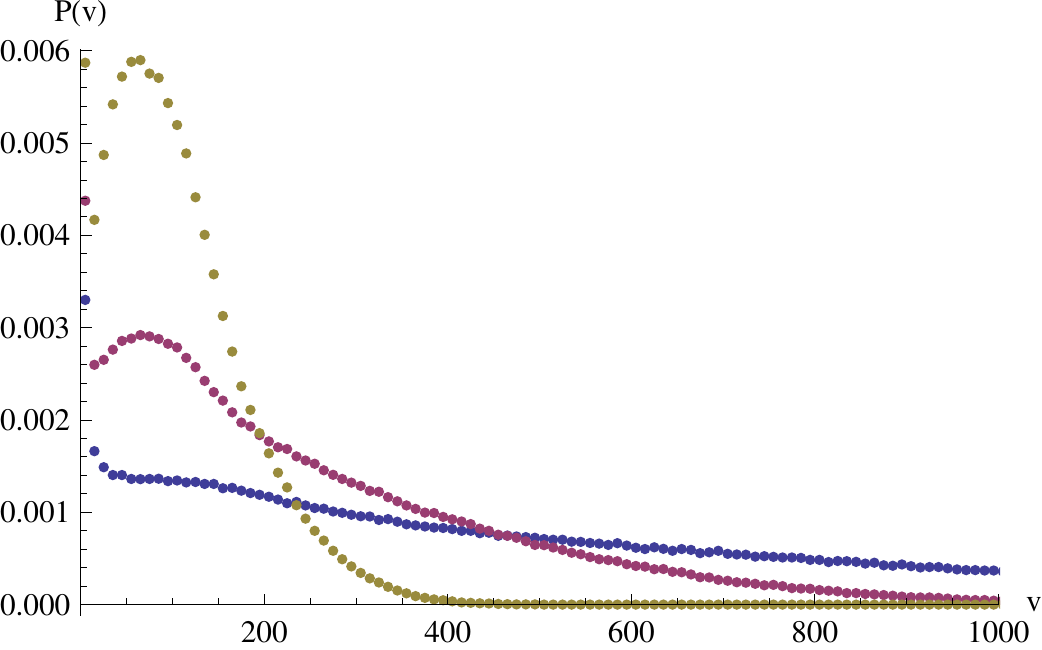}
\caption{The the recoil magnitude distribution for dry mergers
displaying a tail extending beyond $1000\ \KMS$. 
We also display the distributions representing recoils for
wet mergers for
hot and cold accretion disks. The cold disk leads to a recoil
velocity distribution highly peaked 
at around $v\approx80\ \KMS$ while the hot accretion disk extends
the magnitude of the recoil to several hundred $\KMS$. 
}
\label{fig:kickdist}
\end{center}
\end{figure}

The same statistical analysis can be made with the magnitude of
the recoil velocity of the remnant final black hole when we
consider a set of $10^6$ binaries with uniform distributions of
mass ratios in the range $0\leq q\leq 1$. For dry mergers we consider 
uniform orientations of the spin
directions, and uniform distributions of spin magnitudes.
We evaluate Eq.~(\ref{eq:Pempirical}) each time and obtain
the distribution with the extended tail beyond $1000\ \KMS$ in
Fig.\ \ref{fig:kickdist}. The other two distributions correspond
to the wet mergers with $0.3\leq\alpha_i\leq0.9$ and orientations within 10 deg and 30 deg for cold and hot accretion disks respectively
according to Ref.\ \cite{Dotti:2009vz}. The results show a tighter
distribution around low recoil velocities for cold than for hot accretion
disks around the merging black holes.

Finally, another use of the remnant formulae can be found in modeling
waveforms in the intermediate and
small mass ratio limits using the techniques of
Ref.~\cite{Lousto:2010tb} by providing an accurate a priori
estimation for the
background black-hole mass and spin.

\section*{acknowledgments}
We thank E.~Berti for discussion on the resonances of PN evolutions
and M.~Volonteri on describing the SPH results in detail.
We gratefully acknowledge NSF for financial
support from grant PHY-0722315, PHY-0653303, PHY 0714388, and PHY
0722703; and NASA for financial support from grant NASA
07-ATFP07-0158 and HST-AR-11763.  

\section*{References}
\bibliographystyle{iopart-num}
\bibliography{../../../Bibtex/references}
\end{document}